\documentstyle[float,prl,aps]{revtex}

\begin{document}
\draft
\preprint{LA-UR 93-4099}

%% %% \title{Comment on Tuning of the Charge-Density-Wave in the
%% %% Halogen-Bridged Transition-Metal Linear-Chain Compounds }
%%
%% \title{Comment on Tuning of the CDW in MX Chains}
%% \author{J.~Tinka Gammel}
%% \address{Theoretical Division,
%% Los Alamos National Laboratory, Los Alamos, New Mexico 87545 }
%% \date{10 November 1993}
%% \maketitle
%% \pacs{1992 PACS:
%% 71.38.+i, %% pol and el-ph
%% 71.25.Tn, %% band structure
%% 71.20.Hk  %% elec dos in polymers etc
%% }

\narrowtext
%\widetext
\twocolumn
\centerline{\bf Comment on Tuning of the CDW in MX Chains}
\vskip 12truept

\begin{floating}[b] \vskip 3.25truecm
\vskip 60truept
\vbox{\widetext
\begin{figure}\caption{\label{fig_1}
Quadratic (solid, Eq.~{\protect\ref{quadratic}}) and quartic (dashed,
Eq.~{\protect\ref{quartic}}) fits to the lattice energy (o) reported
in  Ref.~{\protect\cite{mebprl}} (a) as a function of $\delta$ for (top
to bottom) $d_{MM}$= 5.18, 5.29, 5.37, 5.45, and 5.55 \AA\ and
(b) as a function of $d_{MM}$ for $\delta$=0
%% for $\delta/d_{MM}$= 0.00, 0.01, 0.02, 0.03, and 0.04.
with the Ref.~{\protect\cite{mebprl}} fit shown ($\times$).
(c) Same as (a) but for the total energy.  Note the fit to
quadratic $M$$X$ springs is good, and the quartic fit excellent.
}\end{figure}
\narrowtext}
\vskip -60truept
\end{floating}

In Ref.~\cite{mebprl}, the authors performed
local-density-approximation (LDA) calculations and fit to the two-band
model  \cite{longmx} we proposed for halogen-bridged transition-metal
linear-chain compounds ($M$$X$ chains).  Based on that fit, the authors
argue a hard-core repulsive term must be added to the lattice
potential. However,
%% the authors fit only to the variation in total energy as a function
%% of percent dimerization for a fixed $M$$M$ distance  and did not fit
%% the variation in total energy with $M$$M$ distance \cite{meb}. Thus
the fit was incorrectly calculated \cite{meb}.
The LDA total energy surface is {\it not} reproduced %% correctly
by the parameters listed in Ref.~\cite{mebprl}, and the conclusion
as to the importance of anharmonic corrections must be reexamined.

As our two-band Hamiltonian with the PtBr parameters
reported in Ref.~\cite{mebprl} ($t_0$=1.466, $\alpha$=1.15, $e_0$=2.37,
$\beta_M$=$\beta_X$=0) reproduces the LDA electronic bands over
the distortion range studied \cite{hesitate} --
a remarkable success for our simple two-band model --
%% (but see below) -- using Mathematica
I fit the lattice energy (LDA total energy from Ref.~\cite{mebprl}
minus the calculated electronic energy).  A fit to the form of the
potential in Ref.~\cite{mebprl} which {\it correctly} reproduces the
%% LDA result is \begin{eqnarray}
%% E_{lat}^{[1]}& =&
%%   -0.168 + 304.9{\rm a}^2 + 2814.3{\rm a}^3 + 7817.7{\rm a}^4
%%     \label{mebform} \\ &&
%%    + 7.65({\rm a}^2 + \delta^2)
%%    + 7.49\big(({\rm a} - \delta)^4 + ({\rm a} + \delta)^4\big) \nonumber\\
%%   && + 0.054\bigg({1\over{(0.8015 + {\rm a} - \delta)^3}}
%%                + {1\over{(0.8015 + {\rm a} + \delta)^3}}\bigg)
%%   \nonumber
%% \end{eqnarray}
%% These values do reproduce the energy surface unlike the values
%% tabulated in Ref.~\cite{mebprl}, but
LDA result can be found, but the energy surface
%% in Fig.~\ref{fig_1}
does not justify this detailed a parameterization.  Fig.~\ref{fig_1}
shows good fits are obtained with quadratic and quartic $M$$X$ springs
\cite{caveat}:
\begin{eqnarray}
E_{lat}^{quadratic} &=&0.004 - 10.8{\rm a} + 10.6({\rm a}^2 + \delta^2)
\label{quadratic}\\
E_{lat}^{quartic\ \ }
   &=&  0.058 -10.3{\rm a} +8.7({\rm a}^2 +\delta^2) \nonumber\\
   & & -5.8({\rm a}^3 +3{\rm a}\delta^2)
       +34.({\rm a}^4 +6{\rm a}^2\delta^2 +\delta^4) ~
\label{quartic}
\end{eqnarray}
in units of [eV/4-sites]
with $\delta$=($d_{MX}^{long}$$-$$d_{MX}^{short}$)/2,
${\rm a}$=($d_{MM}$$-$$d_{MM}^{expt}$)/2, and $d_{MM}^{expt}$=5.55 [\AA].
The dependence of the lattice energy on $M$$M$ distance $d_{MM}$
at $\delta$=0 is quite linear and the quadratic approximation to
the lattice energy surprisingly good. %%  for small distortions.  However,
As noted in Ref.~\cite{mebprl}, Fig.~\ref{fig_1}(c) shows that upon
application of uniaxial pressure the loss of dimerization in the
quadratic approximation is not as rapid as found in LDA.  In contrast
to the assertion in Ref.~\cite{mebprl}, Fig.~\ref{fig_1} clearly shows
a small quartic correction {\it does} reproduce the LDA behavior over
the entire region. However, the magnitude of the lattice parameters and
the relative importance of such quartic terms
%% to model a more rapid loss of dimerization with uniaxial pressure
%% than found in quadratic approximation
is sensitive to the parameterization of the electronic energy.

Finally, %% it is important to note
a $\beta$=0 parameterization of the electronic two-band Hamiltonian
is problematic. For quadratic $M$$X$ springs at $\beta$=0
the CDW ($X$=$M$=$X$--$M$--) and BOW ($X$--$M$=$X$=$M$--)
phases are degenerate \cite{longmx}, in contrast to
the experimental absence of BOW phase compounds.
For the quartic approximation (Eq.~\ref{quartic})
%%, or the parameterization in Ref.~\cite{mebprl},  %% (Eq.~\ref{mebform}),
%%%% this was not the case, but maybe I had an error as I thought it was
the experimental ground state phase, CDW, is not the minimum energy
distortion: the ground state
is the 4-fold degenerate MIX ($X$=$M$--$X$$\cdot\cdot\cdot$$M$--)
phase, where defects show charge/spin separation and other unusual
behavior \cite{mix}.
%% as discussed in Ref.~\cite{mix} in a different parameter region.
%% or in Ref.~\cite{Kenji} in 2D.
While this interesting behavior might occur in Ni-based $M$$X$ chains,
%% the experimental evidence indicates
it does not occur in the Pt-based $M$$X$ chains
studied in Ref.~\cite{mebprl}, which
shows the physical importance of keeping either a $\beta$ term
or an analogous distance dependence of the electron-electron
repulsion as discussed in Ref.~\cite{ivo}.

\vskip 12truept
\noindent{J.~Tinka Gammel}\hfil\break
\hbox{\ \ \ }{\small Theoretical Division, Los Alamos National Laboratory}
\hbox{\ \ \ }{\small Los Alamos, New Mexico 87545 }
\vskip 12truept
\noindent{\small Received 10 November 1993}


\begin{references}
\bibitem{hesitate}
Though the fit is good, to reproduce the bands to the same accuracy
as the quartic lattice correction,
non\-linear electron-phonon coupling should also be included.
A more realistic model in good agreement with experimental data for
strongly distorted systems (PtCl chains), which reduces to our
model for small distortions, is given in Ref.~\cite{ivo}.
\end{references}
\end{document}